\documentclass[aps, prb, superscriptaddress, notitlepage, showpacs, floatfix, twocolumn, tightenlines]{revtex4}

\usepackage{amsfonts, amsmath, amssymb}
\usepackage{amscd, latexsym}
\usepackage{mathrsfs}
\usepackage{graphicx} 
\usepackage{epstopdf} 
\usepackage{dcolumn}
\usepackage{bm}
\usepackage{color}

\begin{document}

\title{Enhancement of Resonant Energy Transfer Due to Evanescent-wave from the Metal}

\author{Amrit Poudel}
\affiliation{\mbox{Department of Chemistry, Northwestern University, 2145 Sheridan Road, Evanston, IL 60208, USA}}

\author{Xin Chen}
\affiliation{\mbox{Center of Nanomaterials for Renewable Energy, State Key Laboratory of Electrical Insulation and Power Equipment,}\\
\mbox{School of Electrical Engineering, Xi'an Jiaotong University, Xi'an, China}}

\author{Mark A. Ratner}
\affiliation{\mbox{Department of Chemistry, Northwestern University, 2145 Sheridan Road, Evanston, IL 60208, USA}}

\begin{abstract}
The high density of evanescent modes in the vicinity of a metal leads to enhancement of the near-field F\"{o}rster resonant energy transfer (FRET) rate. We present a classical approach to calculate the FRET rate based on the dyadic Green's function of an arbitrary dielectric environment, and consider non-local limit of material permittivity in case of metallic halfspace and thin film. In a dimer system, we find that the FRET rate is enhanced due to shared evanescent photon modes bridging a donor and an acceptor. Furthermore, a general expression for the FRET rate for multimer systems is  derived. The presence of a dielectric environment and the path interference effect enhance the transfer rate, depending on the combination of distance and geometry.
\end{abstract}

\maketitle
\newpage

\section{Introduction}

The short range energy transfer phenomena find applications in wide areas of physics and chemistry from mesoscopic to molecular systems,~\cite{Li15, Blum12,Geddes02, Bergfield13} and is critical to our understanding of physical, chemical and biological processes at the molecular level.~\cite{Moerner03} Understanding the physics of an exciton transfer from a donor to an acceptor is also crucial for designing better solar cells and photo-voltaic materials. This has fueled great interest in the study of the effect of electromagnetic environment on the resonant energy transfer, which offers the possibility of controlling and enhancing the energy transfer in molecular systems.~\cite{Li15, Lunz11, Ghenuche14,  Rabouw14} Therefore, obtaining  an  exact  classical  description of the resonant energy transfer under the influence of an electromagnetic environment is imperative to improve the  energy transfer mechanism in molecular systems.

The  interaction  between a dipole  emitter  and its  mirror image under the metallic surface, or equivalently, the scattered evanescent waves from the metal, has  been  studied in varieties of nanoscale systems.~\cite{Li15, Blum12,Geddes02} Furthermore, the  effect  of  metal on the  luminescent  lifetime of a molecule has been known,~\cite{Chance75} and recently, its effect on the relaxation of the semiconducting quantum dot has been studied.~\cite{Poudel13} The presence of metallic environment in the vicinity of the donor-acceptor system introduces a fluctuating electromagnetic field which can modify the donor's emission and acceptor's absorption spectra, as well as the coupling strength between the donor and the acceptor molecules, which can modify the F\"{o}rster resonant energy transfer (FRET) rate.~\cite{Chen10, Beljonne09}  As a result, one can control the transfer rate by tuning the dielectric environment.

The derivation of the classical energy transfer rate using the dyadic Green's function in the local limit of the metallic environment has  been considered before~\cite{Dung02}.  Here we reconsider the derivation based on photon Green's function, connecting it with previous derivation based on molecular polarizability. We also consider the non-local limit of the material permittivity which is an accurate limit at small separation of the donor-acceptor system from the metallic surfaces.  Furthermore, we generalize our derivation to include energy transfer processes in multimer (multi-donor and acceptor) systems in the presence of an arbitrary electromagnetic environment. We emphasize that the usage of the photons' Green function allows us to treat arbitrary photonics/dielectric environment as well as multimer systems efficiently, from both computational and theoretical perspectives.

Two key factors in the FRET rate are: the electronic coupling of the donor and acceptor dipoles and the overlap of the emission and absorption spectra of donor and acceptor molecules. The FRET is considered as an incoherent energy transfer mechanism in which greater overlap between the emission and absorption spectra due to  incoherent broadening can enhance the transfer rate. An incoherent evanescent wave in the vicinity of the metallic environment can participate in the FRET by bridging the emission and absorption spectra due to the shared modes of evanescent waves from the metal. The similar concept of shared phonon modes has been studied previously~\cite{Beljonne09, Hennebicq09} in the context of the enhancement of the FRET rate and recent experiments have shown longer coherence time and enhancement of the exciton transfer in light harvesting proteins.~\cite{Hayes13} In the same spirit of the shared mode approach, the mirror image of the donor dipole can couple to the donor and acceptor at the same time through the evanescent waves, leading to enhanced transfer rate. In addition, in the multimer systems, the shared modes will generate additional pathways for the energy transfer and their interference effect can suppress or enhance the resonant energy transfer rate.

\section{Classical Model and Photon Green's Function}\label{model}

From the classical electromagnetic (EM) theory, the energy flux density of the EM field from the donor to acceptor is given by the  Poynting vector, $\langle \mathbf{S} \rangle_p=\langle \mathbf{E} \times \mathbf{H}\rangle_p$. Adopting the classical perspective by Kuhn and Silbey,~\cite{Kuhn70,  Chance75, Zimanyi10} the energy transfer from donor to acceptor can be understood as two coupled oscillating dipoles (donor and acceptor). The donor radiates an electric field that permeates the acceptor and the acceptor as an antenna receives the energy from this field. Due to the existence of an electromagnetically active environment like metal, the electric field experienced by the donor and the acceptor  will be modified and can be determined by the boundary conditions, $\it{i.e.}$ the geometry of the metallic environment. Here we employ the dyadic Green's function to calculate the modified electric field of a donor molecule due to the presence of metallic environment, the effect of which is fully captured by the relative permittivity $\varepsilon(\vec{r}, \omega)$ of the environment. 

The retarded photon Green's function $\hat{D}$  satisfies the following wave equation~\cite{Lifshitz80}:
\begin{align}
&\Big[\partial_{i}\partial_{j} -\delta_{ij}\Big( \nabla^{2}+\frac{\omega^{2}\varepsilon(
\vec{r}, \omega)  }{c^{2}}\Big)\Big]
D_{ik}(\vec{r}_D,\vec{r}_A, \omega)  \nonumber \\
&= \delta^{3}\left(\vec{r}_D-\vec{r}_A\right)  \delta_{jk}\,.
\end{align}
Here $c$ is the speed of light in vacuum and $\vec{r}_D$ and $\vec{r}_A$ are positions of donor and acceptor, respectively.  We briefly pause here to mention that the dyadic Green's function contains two important points of information. First, it provides the electric field $\vec{E}_D$ of the donor dipole $\mu^{elec}_D \vec{n}^{elec}_D$, where $\mu^{elec}_D(\omega)$ is the strength or magnitude of the electric dipole and $\vec{n}^{elec}$ is the unit dipole, located at $\vec{r}_D$ in the presence of arbitrary metallic environment,
\begin{align}\label{DGF2EM}
\vec{E}_D(\vec{r},\vec{r}_D,  \omega) = \frac{\omega^2 \mu^{elec}_D(\omega)}{c^2 \varepsilon_0} \mbox{Re}[\hat{D}(\vec{r}, \vec{r}_D,  \omega)] \cdot \vec{n}^{elec}_D\, .
\end{align}
For molecules with finite volumes that cannot be approximated by point dipoles, the dipole electric field is obtained by integrating the dyadic Green's function over the volume of the molecule.  
Second, it provides the thermal correlation of the photon reservoir, which includes both evanescent and propagating waves, governed by the spectral density $S_{ij}$
\begin{align}
S_{i,j}(\vec{r}_D, \omega) = \int_{-\infty}^{+\infty} d\tau \langle E_i(\vec{r}_D, 0) E_j(\vec{r}_D, \tau) \rangle e^{-i \omega \tau} \,,
\end{align}
which is related to the dyadic Green's function via fluctuation-dissipation theorem:~\cite{Joulian05, Lifshitz80}
\begin{align}
S_{ij}(\vec{r}_D, \pm\omega) = \frac{\hbar \omega^2}{\epsilon_0 c^2} \mbox{Im} [D_{ij}(\vec{r}_D, |\omega|)] [1 + N(\pm\omega)]\,,
\end{align}
where $N(\omega)$ is the Planck's function and the indices $i,j$ run over Cartesian coordinate axes $\hat{x}, \hat{y}$ and $\hat{z}$.

\subsection{Transfer Rate in Dimer Systems}
With this information, we are now in a position to  express the rate of energy transfer in terms of Green's function. We start from  the previously known expression of the energy transfer rate per unit volume in the time domain:~\cite{Zimanyi10, Duque15}
\begin{equation}
\dot{Q}(t) = \vec{E}_D^\dag(\vec{r}_A,\vec{r}_D,  t)\cdot \dot{\vec{P}}_A(t),
\end{equation}
where $\vec{P}_A(t)$ is  the polarization of an acceptor in the time domain and $ \vec{E}_D(\vec{r}_A,\vec{r}_D,  t)$ is the electric field due to the donor molecule at the acceptor's location $\vec{r}_A$. To compare with the FRET rate, we transform $\dot{Q}$ into the frequency domain,
\begin{align}
\tilde{\dot{Q}}(\omega) = -i  \int_{-\infty}^{\infty} d\omega'\, \omega'\, \vec{E}_D^\dag(\vec{r}_A,\vec{r}_D,  \omega-\omega')\cdot \vec{P}_A(\omega'),
\end{align}
Using the acceptor molecule's polarization $\vec{P}_A(\omega) = \varepsilon_0 \hat{\chi}_A(\omega) \cdot \vec{E}_D(\vec{r}_A, \vec{r}_D, \omega)$, where $\hat{\chi}_A(\omega)$ is the polarizability tensor of the acceptor, the energy transfer rate becomes:
\begin{align}
\tilde{\dot{Q}}(0) =-i \varepsilon_0 \int_{-\infty}^{\infty} d\omega\, \omega \, &\vec{E}_D^\dag(\vec{r}_A,\vec{r}_D,  -\omega) \cdot \hat{\chi}_A(\omega) \cdot \nonumber \\
 &\vec{E}_D(\vec{r}_A, \vec{r}_D, \omega)\,,
\end{align}
which in time-domain reduces to the previous result derived by Chance $et.\,al$~\cite{Chance75}. Using Eq.~\ref{DGF2EM} together with $\hat{\chi}_A(\omega) = \frac{1}{V \varepsilon_0\hbar}\chi_A(\omega) \vec{n}^{elec}_A \vec{n}^{elec}_A$, where $\vec{n}^{elec}$ is the unit electric dipole moment of the acceptor molecule and $V$ is the total volume, and dividing the energy transfer rate by $\hbar \omega$, we obtain the rate of resonant energy transfer:
\begin{align}\label{rate}
\gamma_{FRET} = \int^{\infty}_{-\infty} d\omega\,  &\frac{\omega^4}{\hbar ^2 \varepsilon_0^2 c^4}\, \sigma_A(\omega)\, \sigma_D(\omega)\, \times \nonumber \\
& |\vec{n}^{\dag elec}_A \cdot \hat{D}(\vec{r}_A, \vec{r}_D, \omega) \cdot \vec{n}^{elec}_D|^2\,.
\end{align}
Here we defined the absorption spectrum of the acceptor molecule $\sigma_A (\omega) = \mbox{Im}[\chi_A(\omega)]$ and the donor emission spectrum $\sigma_D$, which according to Fermi's Golden rule, is $\sigma_D(\omega)= \mu^{elec}_D(\omega)^2\int^{\infty}_{-\infty} d\omega_\alpha F^2(\omega_{\alpha}, \omega_{\alpha}+\omega) g(\omega_{\alpha})$, where $F$ is the Franck-Condon factor and $g(\omega_{\alpha})$ is the distribution function of the donor state $\alpha$ such that $\int d\omega_{\alpha}\,g(\omega_{\alpha}) = 1$. In our model, the effect of electromagnetic environment is fully captured by the photons Green's function, which can be obtained computationally by solving Maxwell's equation.

Eq.~\ref{rate} shows that the transfer rate is determined by the overlap of the donor emission spectrum $\sigma_A(\omega)$, acceptor absorption spectrum $\sigma_D(\omega)$ and the dyadic Green's function $ \hat{D}(\vec{r}_A, \vec{r}_D, \omega)$. These three terms represent two fluctuation sources--phonon bath (linewidth broadening of emission and absorption spectra) and the evanescent component of the fluctuating  electromagnetic waves.

\subsection{Transfer Rate in Multimer Systems}
Using the dyadic Green's function, we extend the FRET rate from dimer to multimer systems. Assuming that the system consists of $D_l$ with $l={1,\cdots,N_D}$ donors and $A_m$ with $m={1,\cdots,N_A}$ acceptors, one can write the dyadic Green's function for all possible interaction between the donors and acceptors using the same photon Green's function described above, except with different labels of $\vec{r}_{D_l}$ for a donor molecule location and $\vec{r}_{A_m}$ for the location of an acceptor molecule. The multimer FRET (MFRET) rate is given by
\begin{align}
\label{QMC}
\dot{Q}(t) = \sum_{l=1}^{N_D} \sum_{m=1}^{N_A} \vec{E}^\dag(\vec{r}_{A_m}, \vec{r}_{D_l}, t)\cdot \dot{\vec{P}}_{A_m}(t)
\end{align}
The electric field experienced by the $m$-th acceptor  due to the $l$-th donor consists of a direct field originating from the $l$-th donor  and an indirect field mediated by other donors and acceptors, due to their induced dipole moments by the $l$-th donor field:
\begin{align}
\label{EDMC}
&\vec{E}(\vec{r}_{A_m},  \vec{r}_{D_l}, \omega) =  \frac{\mu^{elec}_{D_l}(\omega)\,\omega^2}{\varepsilon_0c^2}\Bigg[\hat{D}_{A_m, D_l}(\omega) \cdot \vec{n}^{elec}_{D_l} \\
&+  \frac{V \omega^2}{c^2} \Bigg \{ \sum_{l'\neq l}^{N_D} \hat{D}_{A_m, D_{l'}}(\omega) \cdot  \hat{\chi}_{D_{l'}} (\omega) \cdot \hat{D}_{D_{l'}, D_l}(\omega)\cdot \vec{n}^{elec}_{D_l} \nonumber \\
&+ \sum_{m'\neq m} ^{N_A} \hat{D}_{A_m, A_{m'}} (\omega) \cdot \hat{\chi}_{A_{m'}} (\omega)  \cdot \hat{D}_{A_{m'}, D_l}(\omega) \cdot \vec{n}^{elec}_{D_l} \Bigg \} \Bigg]\nonumber
\end{align}
Here we considered the first order indirect field only. However, we stress that our formalism can be generalized to accommodate higher order indirect fields. The induced polarization of the $m$-th acceptor $\vec{P}_{A_m}$ is given by $\vec{P}_{A_m} = \sum_{j=1}^{N_D}\varepsilon_0 \hat{\chi}_{A_m}(\omega) \cdot \vec{E}(\vec{r}_{A_m}, \vec{r}_{D_j}, \omega) +   \sum_{i\neq m}^{N_A}\varepsilon_0 \hat{\chi}_{A_m}(\omega) \cdot \vec{E}(\vec{r}_{A_m}, \vec{r}_{A_i}, \omega)$, where $\vec{E}(\vec{r}_{A_m}, \vec{r}_{A_i}, \omega) =  \frac{\mu^{elec}_{A_i}(\omega)\,\omega^2}{\varepsilon_0c^2}\hat{D}_{A_m, A_i}(\omega) \cdot \vec{n}^{elec}_{A_i}$. Similar to our previous derivation of the transfer rate in a dimer system,  we first transform Eq.~\ref{QMC} into the frequency domain and then use Eq.~\ref{EDMC}, along with the acceptor's polarization,  to obtain the rate of resonant energy transfer in multimer systems:
\begin{align}\label{mcfret}
&\gamma_{MFRET} = \sum_{m=1}^{N_A}  \sum_{l, j=1}^{N_D} \int^{\infty}_{-\infty} d\omega\,  \frac{\omega^4(\omega)}{\hbar ^2 \varepsilon_0^2 c^4}\, \Bigg[\sigma_{A_m}(\omega)\, \sigma_{D_l, D_j} (\omega) \times \nonumber \\
& [\vec{n}^{\dag elec}_{A_m} \cdot \hat{D}_{{A_m}, {D_l}} (\omega) \cdot \vec{n}^{elec}_{D_l}]^\dag[\vec{n}^{\dag elec}_{A_m} \cdot \hat{D}_{{A_m}, {D_j}} (\omega) \cdot \vec{n}^{elec}_{D_j}] 
\end{align}
Here we considered terms that are of the order $\mathcal{O}(\chi(\omega))$ only and ignored all other higher order terms. Eq.~\ref{mcfret} is equivalent to the previously derived result~\cite{Duque15}, where neither dielectric environment was  considered nor Green's function formalism was employed. We emphasize that our formalism readily allows us to include other higher order terms in the transfer rate. The higher order terms associated with the indirect field (the first order indirect field is presented in Eq.~\ref{EDMC}) can potentially induce stronger path interference in multimer system of certain geometrical arrangement. Furthermore, Eq.~\ref{mcfret}  includes more  channels/paths for energy transfer from donors to acceptors. The interplay among different paths, phonon and photon baths can lead to enhancement or reduction of the transfer rate in such systems.

\section{Rate Enhancement by Evanescent Waves}\label{enhance}
We start by considering a dimer donor-acceptor system to understand the distance dependent enhancement due to coupling of the dimer with the evanescent wave from the metallic surface using bulk material permittivity in the local limit. We assume that the separation distance between the donor and the acceptor $\vec{R} = \vec{r}_D -\vec{r}_A$ is smaller than the vacuum wavelength $\lambda\equiv c/\omega$, that is, $|\vec{R}|/\lambda <<1$, and  neglect any retardation effects. We consider silver metal with plasma frequency $\omega_p = 1.45\times10^{16}$ rad/s, and the electron scattering rate $\nu=2.83\times10^{13}$ rad/s.~\cite{Ordal85} The emission and absorption spectra of donor and acceptor molecules in vacuum are assumed to be Gaussian functions centered around $\omega_d=1.96\,eV$, and $\omega_a=3.14\, eV$, respectively, with a standard deviation of one-fifth of the donor's emission frequency.~\cite{Jang13} Without loss of generality, we assume that both dipoles point in the $z$-direction. We  then compare the energy transfer rate from donor to acceptor in the presence and absence of metal. The analytical expression of Green's functions for planar geometries are presented in the App.~\ref{app1}.
\begin{figure}
 \includegraphics[width=0.8\columnwidth]{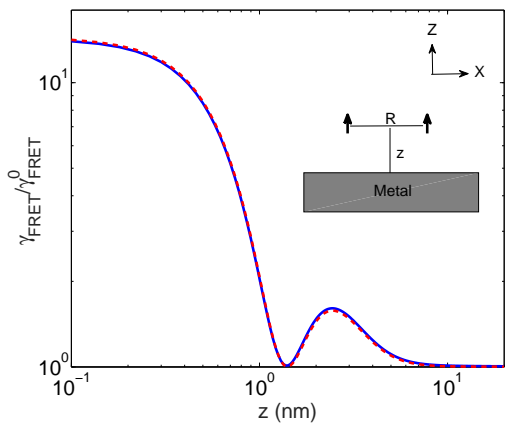}
 \caption{(Color online) Ratio of the  energy transfer rate $\gamma_{FRET}$ from donor to acceptor in the presence of silver half-space (dashed red) and silver thin film (solid blue) to the rate of energy transfer in vacuum $\gamma^0_{FRET}$ as a function of distance $z_D=z_A$ from the metallic surface. The donor-acceptor separation distance $R = |x_A - x_D|=4\, nm$ and thin film thickness $a = 10\, nm$.}
 \label{f1}
\end{figure}

In Fig.~\ref{f1} we plot the ratio of donor-acceptor energy transfer rate in the presence of metallic half-space and thin film to the rate of energy transfer in vacuum. Here, we assume a donor and an acceptor are located at a distance $z_D =z_A$ above the silver surface. We find that the energy transfer rate is modified significantly in the presence of metallic environment at small separation distances from the metal.  For distances $z_D=z_A$ shorter than the donor-acceptor separations, $R\equiv|x_A - x_D|$,  the energy transfer rate is strongly enhanced in the presence of metal. In the opposite limit, i.e., $z_D=z_A> R$, the transfer rate mediated by metallic environment is suppressed and it  saturates to the vacuum transfer rate. At a small separation distance from the metal, there exists a high density of evanescent modes which is responsible for enhancement of the transfer rate in the presence of metal. Since these modes decay exponentially with distance, they have less effect on the transfer rate as we move further away from the metal.

Fig.~\ref{f1} also reveals an interesting path interference effect in the resonant energy transfer process. In the presence of metallic environment, excitons have two pathways$--$direct way from the donor to the acceptor and indirect way mediated by metal$--$to transfer from the donor to the acceptor. The destructive interference between these pathways leads to suppression of energy transfer around $z_D=z_A \approx 1\, nm$ and the constructive interference leads to enhancement of the transfer rate around $z_D=z_A \approx 3\,nm$.

\begin{figure}
 \includegraphics[width=1.0\columnwidth]{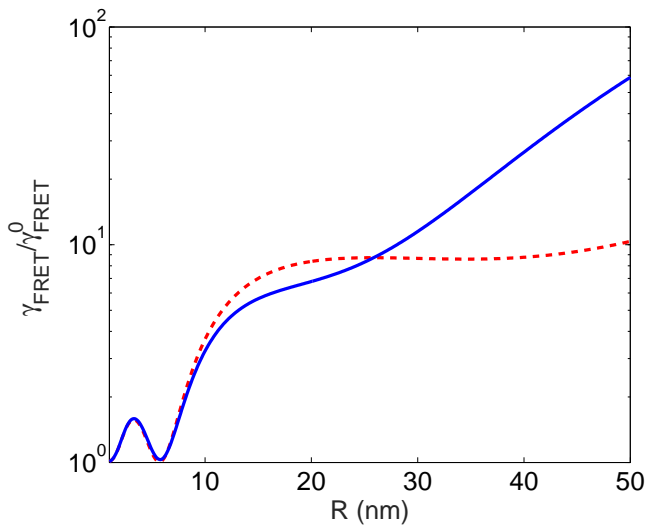}
 \caption{(Color online) Ratio of the  energy transfer rate $\gamma_{FRET}$ from a donor to an acceptor in the presence of silver half-space (dashed red) and thin film (solid blue) to the rate of energy transfer in vacuum $\gamma^0_{FRET}$ as a function of donor-acceptor separation distance $R \equiv |x_A - x_D|$. The  distance from the metallic surface is held constant at $z_D = z_A = 2\, nm$ and the thin film thickness $a = 10\,nm$.}
 \label{f2}
\end{figure}

Next, we vary the separation distance $R$ between the donor-acceptor molecules keeping the distance $z_D = z_A$ from the metallic surface fixed. The result is plotted in Fig.~\ref{f2}. The metallic half-space and thin film result deviates significantly for separation $R$ greater than film thickness $a$. As the separation gets larger, the evanescent modes on the opposite face of the thin film also contribute to the coupling of donor-acceptor system, leading to enhancement of the transfer rate.

\begin{figure}
 \includegraphics[width=1.0\columnwidth]{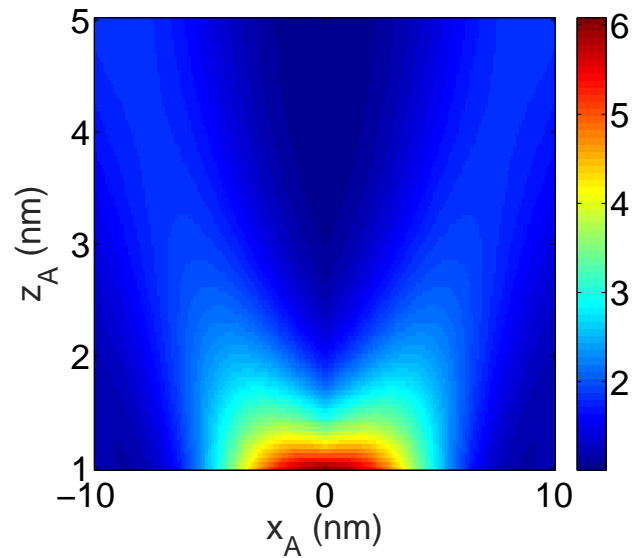}
 \caption{(Color online) Ratio of the  energy transfer rate $\gamma_{FRET}$ from a donor to an acceptor in the presence of thin silver film of thickness $a = 10\,nm$ to the rate of energy transfer in the vacuum $\gamma^0_{FRET}$ as a function of acceptor position $x_A$ and $z_A$, keeping the donor position fixed at $x_D = 0$ and $z_D = 5\,nm$ from the silver surface.}
 \label{f3}
\end{figure}

In Fig.~\ref{f3}, we fix the position of the donor at $z_D=5\,nm$ above the silver surface and vary the position of the acceptor in  both $x$- and $z$- directions. We find that the energy transfer rate is significantly modified when the acceptor is close to the silver surface, as shown in Fig.~\ref{f3}. Near the metal surface, the presence of high density of evanescent modes significantly increase the transfer rate.

\begin{figure} 
 \includegraphics[width=1.0\columnwidth]{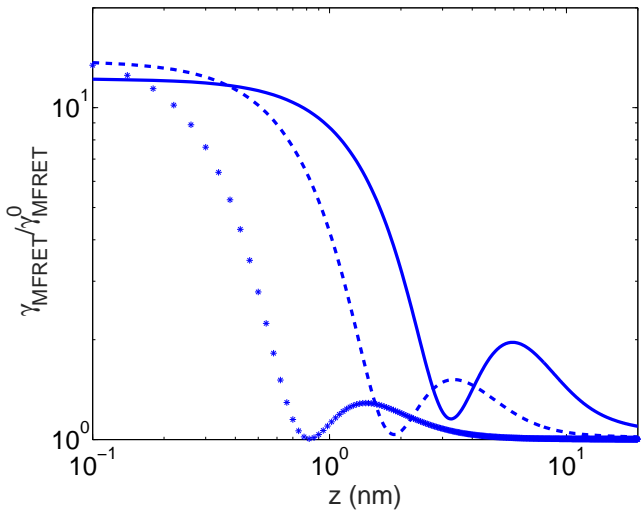}
 \caption{(Color online) Ratio of the  energy transfer rate $\gamma_{MFRET}$  in a multimer system of a linear chain of two (solid blue) three (dashed blue) and six (dashed blue) dipoles pointing in the z-direction in the presence of thin silver film of thickness $a=10\,nm$ to the rate of energy transfer in the vacuum $\gamma^0_{MFRET}$ as a function of distance $z_D=z_A$ from the metallic surface. The exciton transfers from the left-most donor to the right-most acceptor  which are separated by the distance $R = |x_A - x_D|=10\, nm$.}
 \label{f4}
\end{figure}

Next, we investigate the effect of the dielectric environment  on the transfer rate of a multimer (multi-chromophore) system. We consider a linear chain of donor and acceptor molecules that are placed horizontally above the silver surface. We assume that the exciton transfers from the left most donor to the right most acceptor in the presence of intermediate donor molecules in a linear chain of the form $\,D\,D\, D\, \ldots \, \ldots\, A$, where $D$ and $A$ represent donor and acceptor molecules, respectively. The ratio of multimer FRET rate is plotted in Fig.~\ref{f4}. We find that the point where destructive interference occurs, shifts from a larger to a smaller distance from the metal surface as we increase the number of intermediate donor molecules while keeping the length of the chain fixed. For such an arrangement, the distance between the donor and the acceptor molecules drops. Consequently, the point where destructive interference occurs also shifts toward a smaller distance from the metal, as seen in Fig.~\ref{f4}.

\begin{figure}
 \includegraphics[width=1.0\columnwidth]{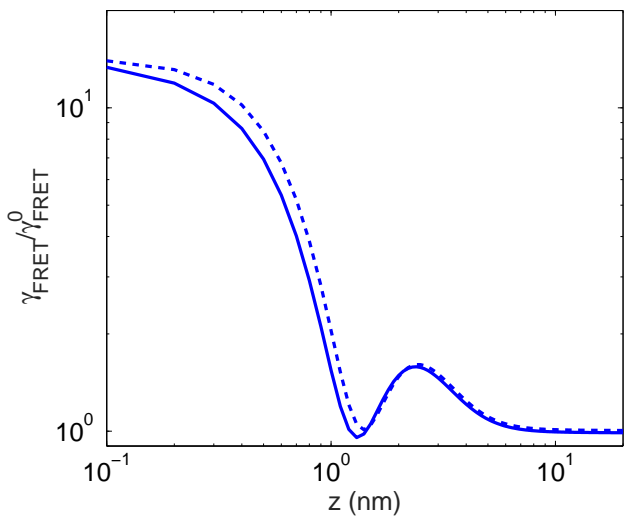}
 \caption{(Color online) Ratio of the  energy transfer rate $\gamma_{FRET}$ from donor to acceptor in the presence of silver thin film of thickness  $a=10 \,nm$ in the local limit (dashed  blue) and non-local limit of permittivity (solid blue) to the rate of energy transfer in vacuum $\gamma^0_{FRET}$ as a function of distance $z_D=z_A$ from the metallic surface. The donor-acceptor separation distance $R = |x_A - x_D|=4\, nm$.}
 \label{f5}
\end{figure}
In the above discussions, we assumed a local limit of the dielectric environment by considering the bulk value of the metal permittivity. However, it is important to emphasize that the bulk permittivity does not necessarily describe the surface effect very accurately, especially when the separation distance between the metal and donor-acceptor system becomes smaller than the mean free path of the electron in the metal.  Since the energy transfer rate is modified only at a small separation from the metal, it is necessary to take an accurate model of the permittivity of the metal to obtain the better estimate of the effect of the metallic environment on the transfer rate. To this end, we calculate the transfer rate using the non-local expression of the permittivity of a metallic thin film presented in the App.~\ref{app2}. The results are plotted in Fig.~\ref{f5}. 

We find that the non-local result does not differ significantly from the  local result, unlike the case of relaxation rate in semiconducting quantum dots, where non-local result deviates significantly from the local result at small distances from the metal.~\cite{Langsjoen12, Kolkowitz15} We attribute this difference in behavior to the fact that in case of energy transfer, Green's function is evaluated at the spatial locations of a donor and an acceptor, which are always separated by a finite distance. However, in  the case of relaxation rate calculation, Green's function is always evaluated at the location of a single molecule or a quantum dot. The presence of finite length scale in the case of energy transfer prohibits contribution from large momentum wavevectors of the evanescent modes in the local limit and does not lead to divergence of the transfer rate even at small separations from the metal. As such, the non-local limit only offers small corrections and the result does not deviate significantly from the local approximation.

\section{Conclusion}\label{conclusion}

We have studied the effect of metallic environment on the resonant  molecular energy transfer using the classical model based on the dyadic Green's function in local and non-local limits of the material permittivity. Our results demonstrate that the evanescent waves from the metallic halfspace and thin film can enhance the FRET rate between the donor and the acceptor since they are simultaneously coupled to the same evanescent mode from the metal. Furthermore, we have extended  our model to multimer systems in arbitrary dielectric environments.  The numerical results  indicate that since all donors and acceptors are under the influence of the evanescent wave, the path interference through all the incoherent channels can enhance or suppress the energy transfer rate with different arrangements of donors and acceptors and their distances from the metal surface.

Our work suggests that by utilizing metallic environments and different arrangements of dipoles/molecules, the resonant energy transfer rate can be modulated.  In the future, we would like to implement computational method based on our theoretical framework to improve F\"{o}ster's energy transfer rate by tailoring the metallic geometry. In addition, further studies to identify optimal geometrical arrangement of donors and acceptors to guide the energy flow in nanoscale systems will find applications in designing nano-materials for efficient harvesting of solar light. Furthermore, the current model can be used, with proper extension, to study heat transfer in nanoscale systems.

\section{Acknowledgement}
A.P. would like to thank Maxim G. Vavilov for fruitful discussions. A.P. and M.R. were supported as part of the ANSER Center, and Energy Frontier Research Center funded by the US  Department of Energy, Office of Science, Office of Basic Energy Sciences, under award no. DE-SC0001059. A.P. and M.R. also partially supported by the Center for Bio-Inspired Energy Science, an Energy Frontier Research Center funded by the U.S. Department of Energy, Office of Science, Basic Energy Sciences under Award DE-SC0000989. X.C. would like to thank Profs. Chunming Niu and Yonghong Cheng for support and encouragement. X.C. has been supported by the start-up funding of Xi'an Jiaotong University and  acknowledges financial support from State Key Laboratory of Electrical Insulation and Power Equipment, School of Electrical Engineering, Xi'an Jiaotong University.


\appendix

\section{Dyadic Green's Function in Local Limit}\label{app1}

In this section, we present a few dyadic Green's function that are analytically tractable.  First, we consider a special case of a homogeneous medium with relative permittivity $\varepsilon(\vec{r}, \omega) = \varepsilon_0$ and $k_{0} = \frac{\sqrt{\varepsilon_0} \omega}{c}$, for which the dyadic Green's function $\hat{D}^0(\vec{R}, \omega)$ is given by:
\begin{align}
\label{D0}
 \hat{D}^0 (\vec{R}, \omega) &= \Big( \hat{I} + \frac{\nabla \nabla}{k_{0}^2}\Big)\frac{e^{i k_{0} R}}{4\pi R}
 = \frac{e^{i k_{0} R}}{4\pi R}\Big[ \Big(1 + \frac{ik_{0} R - 1}{k_{0}^2 R^2}\Big) \hat{I} \nonumber \\
 & + \Big( \frac{3 -3ik_{0}R - k_{0}^2 R^2}{k_{0}^2 R^2} \Big)\frac{\vec{R} \otimes \vec{R}}{R^2}\Big]\,,
 \end{align}
 where $\hat{I}$ is a $3$ by $3$ diagonal matrix and $\vec{R} = \vec{r} - \vec{r}'$. Below we consider other inhomogeneous geometries, like half-space and thin films, for which it is possible to compute the dyadic Green's function analytically. Such geometries are also useful if we limit ourselves to the situation where the separation of the donor-acceptor system from the metal surface is smaller than the radius of curvature of the surface so that the surface can be assumed to be flat. For a source point at $\vec{r}=(x, y, z)$, a field point at $\vec{r}' = (x', y', z')$ and defining a two dimensional vector $\vec{\rho} = (x-x', y-y')$, the scattering Green's function for a half space or a thin film geometry with the material permittivity $\varepsilon_2$ in both local and non-local limits, and the surrounding permittivity $\varepsilon_1$ is given by:
\begin{align}
\label{Dxx_Dzz}
&D_{xx}(\vec{r}, \vec{r}', \omega) =\frac{i c^2}{8 \pi ^2\omega^2} \int_{0}^{2 \pi} d\theta \int_{0}^{\infty}\frac{pdp}{\varepsilon_1 q_1}\,e^{iq_1(z+z') + i\vec{p}\cdot\vec{\rho}}  \nonumber \\
&\times \Big[\frac{\omega^2}{c^2}r_{s}(p) \sin^2\theta -q_1^2r_{p}(p) \cos^2\theta\Big] \,,\\
&D_{zz}(\vec{r}, \vec{r}', \omega)=\frac{i c^2}{8\pi^2\omega^2} \int_{0}^{2\pi} d\theta \int_{0}^{\infty}\frac{p^3}{\varepsilon_1 q_1}dp \nonumber \\
&\times e^{iq_1(z+z') + i \vec{p}\cdot\vec{\rho}} \,\, r_{p}(p) \,,
\end{align}
where $p$ is the transverse  and $q_1$ is the z-component of the wave vector, with $q_1=\sqrt{\omega^2/c^2\varepsilon_1-p^2}$. All other components of the dyadic Green's function can be computed from these components.

In the above expression, $r_s$ and $r_p$ are the Fresnel reflection coefficients of the medium $2$ with respect to the medium $1$. For the half space geometry with the material permittivity of the medium 2 in the local limit,  these coefficients are given by the following expressions:
\begin{align}
\label{hs}
r_s(p) = \frac{q_1 - q_2}{q_1 + q_2 } \quad\mbox{and}\quad r_p(p) = \frac{q_1 \varepsilon_2 - q_2 \varepsilon_1}{q_1 \varepsilon_2 + q_2 \varepsilon_1}\,,
\end{align}
where $q_2 = \sqrt{\omega^2/c^2\varepsilon_2 - p^2} $.

Similarly, for a thin film geometry of thickness $a$ and the local permittivity $\varepsilon_2$ surrounded by another medium of permittivity $\varepsilon_1$ on both sides of the thin film, one can also derive the Fresnel reflection coefficients analytically~\cite{Jones69}. The results are:
\begin{subequations}
\begin{align}
r_s(p) = \frac{q_1^2 - q_2^2}{q_2^2 +q_1^2 + 2i q_1 q_2 \cot(q_2 a)} \\
r_p(p) = \frac{\varepsilon_2^2 q_1^2 - \varepsilon_1^2 q_2^2}{\varepsilon_1^2q_2^2 + \varepsilon_2^2 q_1^2 + 2i \varepsilon_1 \varepsilon_2 q_1 q_2 \cot(q_2 a)}
\end{align}
\end{subequations}

\section{Dyadic Green's Function in Non-Local Limit}\label{app2}

In the non-local limit, when the permittivity of the medium 2 depends on a position, $\varepsilon_2 (\vec{r}, \omega)$, the Fresnel reflection coefficients of the half space geometry take different forms and are given by the following expressions\cite{Ford84}:
\begin{subequations}
\begin{align}
r_s(p) &= \frac{\frac{2ic^2}{\pi \omega^2} \int_0^\infty \frac{d\eta}{\varepsilon^t_2(\vec{k}, \omega) - c^2k^2/\omega^2} - \frac{\varepsilon_1}{q_1}}{\frac{2ic^2}{\pi \omega^2} \int_0^\infty \frac{d\eta}{\varepsilon^t_2(\vec{k}, \omega) - c^2k^2/\omega^2} + \frac{\varepsilon_1}{q_1}}  \\
r_p(p) &=  \frac{\frac{q_1}{\varepsilon_1} - \frac{2i}{\pi} \int_0^\infty \frac{d\eta}{k^2} \{ \frac{\eta^2}{\varepsilon^t_2(\vec{k}, \omega) - c^2k^2/\omega^2} + \frac{p^2}{\varepsilon^l_2(\vec{k}, \omega)} \}} {\frac{q_1}{\varepsilon_1} + \frac{2i}{\pi} \int_0^\infty \frac{d\eta}{k^2} \{ \frac{\eta^2}{\varepsilon^t_2(\vec{k}, \omega) - c^2k^2/\omega^2} + \frac{p^2}{\varepsilon^l_2(\vec{k}, \omega)} \}}
\end{align}
\end{subequations}
where $k^2 = p^2 + \eta^2$ and $\varepsilon^{l,t}_2$ are longitudinal and transverse components of the permittivity given by the following expressions:
\begin{align}
&\epsilon_{l}(k,\omega)=1+\frac{3\omega_{p}^{2}}{k^2v_F^2}\frac{(\omega+i\nu)
f_{l}((\omega+i\nu)/kv_F)}{\omega+i\nu f_{l}((\omega+i\nu)/kv_F)},\\
&\epsilon_t(k,\omega)=1-\frac{\omega_p^2}{\omega(\omega+i\nu)}f_t((\omega+i\nu)/kv_F), \\
&f_{l}(x)=1-\frac{x}{2}\ln(x+1)/(x-1),\\
&f_{t}(x)=\frac{3}{2}x^2-\frac{3}{4}x(x^2-1)\ln(x+1)/(x-1).
\end{align}
Here $\nu$ is the electron collision frequency, $\omega_{p}=(ne^{2}/m\varepsilon_0)^{1/2}$
is the plasma frequency, and $v_{F}$ is the Fermi velocity.

Similarly, for a thin film geometry, the Fresnel coefficients in the non-local limit are given by:
\begin{subequations}
\begin{align}
&r_s(p) = \frac{1}{2} \sum_{i=e,o} \frac{\frac{2ic^2}{a\omega^2} \sum_{n=-\infty}^{\infty} F_s(\vec{k}_i, \omega) - \frac{\varepsilon_1}{q_1}}{\frac{2ic^2}{a\omega^2} \sum_{n=-\infty}^{\infty} F_s(\vec{k}_i, \omega) + \frac{\varepsilon_1}{q_1}} \\
&r_p(p) =  \frac{1}{2} \sum_{i=e,o}\frac{\frac{q_1}{\varepsilon_1} - \frac{2i}{a} \sum_{n=-\infty}^{\infty} F_p(\vec{k}_i, \omega)}{\frac{q_1}{\varepsilon_1} + \frac{2i}{a} \sum_{n=-\infty}^{\infty} F_p(\vec{k}_i, \omega)} \\
&F_p(\vec{k}, \omega) = \frac{1}{k^2} \Big\{\frac{\eta^2}{\varepsilon^t_2(\vec{k}, \omega) - c^2k^2/\omega^2} + \frac{p^2}{\varepsilon^l_2(\vec{k}, \omega)}\Big\} \\
&F_s(\vec{k}, \omega) =\frac{1}{\varepsilon^t_2(\vec{k}, \omega) - c^2k^2/\omega^2} \,.
\end{align}
\end{subequations}
Here $k^2_{e=even, o=odd} = p^2 + \eta^2_{e,o}$ with $\eta_e = 2n \pi/a$ and $\eta_o= (2n+1)\pi/a$.

\end{document}